\documentclass[11pt]{article}
\usepackage[a4paper,left=2.4cm,right=2.4cm,top=3.2cm,bottom=3.2cm]{geometry} 

\RequirePackage[colorlinks=true,linkcolor=blue,urlcolor=blue,citecolor=blue]{hyperref}
\RequirePackage[smalltableaux]{ytableau} 
\RequirePackage{amsmath,amsthm,amssymb,mathrsfs,dsfont,bm,graphicx,tikz,makecell,moresize,mathdots,proof}
\RequirePackage[titletoc,title]{appendix}

\newcommand{\nc}{\newcommand}				\newcommand{\rnc}{\renewcommand}
\nc{\x}{\textnormal} \nc{\n}{\operatorname} \nc{\s}{\mathsf} \nc{\bb}{\mathbb}
\nc{\alp}{\alpha}  \nc{\bt}{\beta}			\nc{\gm}{\gamma}  \nc{\Gm}{\Gamma} \nc{\dt}{\delta}
\nc{\Dt}{\Delta}   \nc{\kp}{\kappa}			\nc{\sg}{\sigma}  \nc{\Sg}{\Sigma} \nc{\tht}{\theta}
\nc{\Tht}{\Theta}  \nc{\ld}{\lambda}		\nc{\Ld}{\Lambda} \nc{\om}{\omega} \nc{\Om}{\Omega}
\nc{\phv}{\varphi} \nc{\epsl}{\varepsilon}	\nc{\thv}{\vartheta}
\nc{\Cal}[1]{\mathcal{#1}} \nc{\fr}[1]{\mathfrak{#1}}
\nc{\Ac}{\Cal{A}} \nc{\Bc}{\Cal{B}} \nc{\Cc}{\Cal{C}} \nc{\Dc}{\Cal{D}} \nc{\Ec}{\Cal{E}}
\nc{\Fc}{\Cal{F}} \nc{\Gc}{\Cal{G}} \nc{\Hc}{\Cal{H}} \nc{\Ic}{\Cal{I}} \nc{\Jc}{\Cal{J}}
\nc{\Kc}{\Cal{K}} \nc{\Lc}{\Cal{L}} \nc{\Mc}{\Cal{M}} \nc{\Nc}{\Cal{N}} \nc{\Oc}{\Cal{O}}
\nc{\Pc}{\Cal{P}} \nc{\Qc}{\Cal{Q}} \nc{\Rc}{\Cal{R}} \nc{\Sc}{\Cal{S}} \nc{\Tc}{\Cal{T}}
\nc{\Uc}{\Cal{U}} \nc{\Vc}{\Cal{V}} \nc{\Wc}{\Cal{W}} \nc{\Xc}{\Cal{X}} \nc{\Yc}{\Cal{Y}}
\nc{\Zc}{\Cal{Z}}
\nc{\im}{\n{im}} \nc{\Tr}{\n{Tr}} \nc{\tr}{\n{tr}} \nc{\rank}{\n{rank}} \nc{\rk}{\n{rk}}
\nc{\Bb}{\bb{B}} \nc{\Cb}{\bb{C}} \nc{\Fb}{\bb{F}} \nc{\Hb}{\bb{H}} \nc{\Nb}{\bb{N}}
\nc{\Rb}{\bb{R}} \nc{\Sb}{\bb{S}} \nc{\Zb}{\bb{Z}}
\rnc{\P}{\bb{P}} \nc{\E}{\n{\bb{E}}} \nc{\Eb}{\mathop{{}\bb{E}}}
\nc{\su}{\fr{su}} \nc{\gf}{\fr{g}}  \nc{\hf}{\fr{h}} 
\nc{\RR}{\mathbf{R}} \nc{\id}{\x{id}} \nc{\SU}{\x{SU}} \nc{\Ux}{\x{U}}
\nc{\poly}{\x{poly}} \nc{\Hom}{\x{Hom}}

\nc{\ot}{\otimes} \nc{\Ot}{\bigotimes} \nc{\Oplus}{\bigoplus}
 \nc{\PT}[1]{^{\s{T}_{\!{#1}}}}
\nc{\dg}{^{\dagger}} \nc{\f}[2]{\frac{#1}{#2}}
\nc{\bra}[1]{\langle{#1}|}
\nc{\ket}[1]{|{#1}\rangle}
\nc{\brak}[1]{\langle{#1}\rangle}			\nc{\ketb}[2]{|{#1}\rangle\!\langle{#2}|}
\nc{\ps}[1]{\left(#1\right)} \nc{\pss}[1]{\left[#1\right]} \nc{\psss}[1]{\left\{#1\right\}}
\nc{\xto}[1]{\xrightarrow{#1}} \nc{\dd}{\,\x{d}} \nc{\ee}{\x{e}} \nc{\ii}{\x{i}} 
\nc{\1}{\mathds{1}}  \nc{\2}{\{0,1\}}		\nc{\lims}{\varlimsup}  \nc{\limi}{\varliminf}
\nc{\ty}[1]{{\x{\tiny $#1$}}}				\nc{\too}{\!\!\to\!\!}
\nc{\mdl}{\models}							\nc{\bmdl}{=\joinrel\mathrel|}

\nc{\be}[2]{\begin{#1}#2\end{#1}}
					\nc{\Thm}[1]{\be{thm}{#1}}
			\nc{\Prop}[1]{\be{prop}{#1}}
			\nc{\Coro}[1]{\be{coro}{#1}}
				\nc{\Lem}[1]{\be{lem}{#1}}
\nc{\Pf}[1]{\be{proof}{#1}}
\nc{\Eq}[1]{\be{equation}{#1}} \nc{\Eqn}[1]{\be{equation*}{#1}}
\nc{\Al}[1]{\be{align}{#1}} \nc{\Aln}[1]{\be{align*}{#1}} \nc{\Als}[1]{\be{align}{\be{split}{#1}}}
\nc{\Item}[1]{\be{itemize}{#1}} \nc{\Enum}[1]{\be{enumerate}{#1}}
\nc{\gray}[1]{{\color{gray}#1}} \nc{\red}[1]{{\color{red}#1}} \nc{\blue}[1]{{\color{blue}#1}}
\nc{\mm}[2]{\ps{\begin{array}{c}{#1}\\{#2}\end{array}}}
\nc{\mmnn}[4]{\ps{\begin{array}{cc}{#1}&{#2}\\{#3}&{#4}\end{array}}}
\nc{\ydiag}[1]{\ytableausetup{notabloids}\ydiagram{#1}}
\nc{\ytab}[1]{\ytableausetup{notabloids}\ytableaushort{#1}}

\newcommand{\vertiii}[1]{{\left\vert\kern-0.25ex\left\vert\kern-0.25ex\left\vert #1 \right\vert\kern-0.25ex\right\vert\kern-0.25ex\right\vert}}

\usepackage[affil-it]{authblk}
\title{Quantum certification of state set and unitary channel}
\author[]{Wei Xie\thanks{\href{mailto:xxieww@ustc.edu.cn}{Email: xxieww@ustc.edu.cn}}}
\affil[]{School of Computer Science and Technology, University of Science and Technology of China, Hefei, China}

\rnc\Affilfont{\itshape\small}
\begin{document}
\date{}
\maketitle 

\begin{abstract}
We study efficient quantum certification algorithms for quantum state set and unitary quantum channel.
We present an algorithm that uses $O(\epsl^{-4}\ln|\Pc|)$ copies of an unknown state to distinguish whether the unknown state is contained in or $\epsl$-far from a finite set $\Pc$ of known states with respect to the trace distance.
This algorithm is more sample-efficient in some settings.
Previous study showed that one can distinguish whether an unknown unitary $U$ is equal to or $\epsl$-far from a known or unknown unitary $V$ in fixed dimension with $O(\epsl^{-2})$ uses of the unitary, in which the Choi state is used and thus an ancilla system is needed. We give an algorithm that distinguishes the two cases with $O(\epsl^{-1})$ uses of the unitary, using much fewer or no ancilla compared with previous results.
\end{abstract} 

\section{Introduction and previous work}\label{sec1}


















In the emerging quantum information technology, a fundamental task in building reliable quantum information processing devices is to obtain parameters of an unknown quantum state or device.
The process is called {\em tomography} if all parameters are required to be known, which has been extensively studied since fifty years ago.
Efficient quantum algorithms based on group representation theory have been proposed in \cite{haah2017sample,o2016efficient,o2017efficient}.
In many scenarios, however, we are concerned only with whether the unknown state or operation satisfies specific property.
For example, to assess the quality of a quantum chip after production, one needs only to check whether the circuit is close to a given unitary transformation, and it is unnecessary to get all parameters about this chip.
This process is called {\em certification}, which usually saves samples and storage space compared with the quantum tomography.
A survey on quantum certification is given in \cite{montanaro2016survey}.

The abstract setting for certification can be described as follows.
Given a known set $\Pc$ and an unknown element $x$, a tester (or an algorithm) $\Tc$ either accepts (i.e., reports $x\in\Pc$) or rejects (i.e., reports $x\notin\Pc_\epsl$) with some probability after measuring $x$, where $\Pc_\epsl:=\{y:\x{dist}(y,\Pc)\le\epsl\}$ with dist denoting some metric.
The tester $\Tc$ is eligible if the following conditions hold:

(1) (Completeness) If $x\in \Pc$, $\Tc$ accepts with probability at least 2/3;

(2) (Soundness) If $x\not\in\Pc_\epsl$, $\Tc$ accepts with probability at most 1/3.

The numbers $2/3,1/3$ have no special meaning and can be replaced by any constants $c>1/2,s<1/2$ respectively due to the confidence amplification by using repeating the test.
If the tester accepts with certainty when $x\in\Pc$, we say the tester has perfect completeness.
Recall that the trace distance between two quantum states is $\x{D}(\rho,\sg):=\f{1}{2}\|\rho-\sg\|_1$, and their fidelity is $\x{F}(\rho,\sg):=\|\sqrt\rho\sqrt\sg\|_1$.
Given a known state $\ket\phv$ and an unknown state $\ket\psi$, the aim is to decide whether $\psi=\phv$, i.e., $\ket\psi=\ee^{\ii\tht}\ket\phv$ for some real $\tht$, or $\x{D}(\psi,\phv)\ge\epsl$.
In this task of certification, $\{\ee^{\ii\tht}\ket\phv:\tht\in\Rb\}$ is the known set $\Pc$ and $\ket\psi$ is the unknown element $x$.
It turns out $O(\epsl^{-2})$ copies of given states are sufficient for this task.
The test is simply to perform the POVM $\{\ketb{\phv}{\phv},\1-\ketb{\phv}{\phv}\}$
on $n$ samples independently, and accept if and only if every outcome corresponds to the former POVM element.
If $\x{D}(\psi,\phv)=\sqrt{1-\brak{\psi,\phv}}\ge\epsl$, then $\brak{\psi,\phv}^n\le(1-\epsl^2)^n\le\f{1}{3}$ and one can take $n=O(\epsl^{-2})$ for small $\epsl$.

In order to test whether a pair of unknown states $\rho$ and $\sg$ on $\Hc$ are equal, the swap test \cite{buhrman2001quantum} is usually used.
The swap test is simply a two-outcome measurement $\{P,Q\}$, where $P$ and $Q$ are projectors onto the symmetric subspace and the antisymmetric subspace of $\Hc^{\ot 2}$ respectively.
The first outcome occurs with probability $\f{1}{2}(1+\tr(\rho\sg))$, and when this is the case, we say the swap test accepts or passes.
By applying group theory, it is shown in \cite{buadescu2019quantum} that $O(d/\epsl^2)$ copies of quantum states suffices to distinguish whether an unknown $\rho$ is equal to some known $\sg$ or $\epsl$-far from $\sg$ in trace distance.
This method is efficient since the tomography needs $\Om(d^2)$ copies of quantum states.

Let $\Pc$ be a finite set of pure states such that $\min_{\phv\ne{\phv'}\in\Pc}\x{D}(\phv,\phv')=:\dt$, and let $\psi$ be an unkown pure state.
Then $O(\max\{\epsl^{-2},\dt^{-2}\}\ln|\Pc|)$ copies suffice to distinguish whether $\psi\in\Pc$ or $\x{D}(\psi,\Pc)\ge\epsl$ \cite{wang2011property}.
Following the method in \cite{buadescu2019quantum} we can use $O(\epsl^{-2}|\Pc|)$ copies of the unknown state $\rho$ to distinguish whether $\rho\in\Pc$ or $\min_{\sg\in\Pc}\|\rho-\sg\|_2\ge\epsl$.
We also show that $O(\epsl^{-4}\ln|\Pc|)$ copies of unknown state $\psi$ suffice to distinguish whether $\psi\in\Pc$ or $\x{D}(\psi,\Pc)\ge\epsl$.
When $\dt$ is small compared with $\epsl$, our method exhibits advantage by using notably less samples.

The certification of unitaries is quite different from that of quantum states in that the quantum unitary certification requires a double optimization, one is the input state for the quantum unitary and the other is the choice of measurement after the action of quantum unitary.
The works in \cite{da2011practical,steffen2012experimental,reich2013optimal} used methods based on Monte Carlo sampling to estimate the fidelity of an unknown gate to a fixed one, and also studied the optimality of estimation strategy.
Given an unknown unitary $U$ and a known or unknown unitary $V$, by using the Choi correspondence between quantum unitaries and states, there exists a tester that distinguishes whether their distance is zero or larger than $\epsl$ with $O(\epsl^{-2})$ uses of unitaries.
By using Schur-Weyl decomposition, we show that for fixed dimension of the unitary, only $O(\epsl^{-1})$ uses of the unitaries suffice to achieve the same goal.
Another advantage of our algorithm needs much fewer or no ancilla system compared with the method using Choi state.

Other properties of unitaries have been studied.
Productness can be tested with $O(\epsl^{-2})$ copies \cite{harrow2013testing} using the procedure first discussed in \cite{mintert2005concurrence} which applies the swap test across each corresponding pair of subsystems of two copies of $\ket\psi$. Testing product unitaries can be reduced to testing product pure states using Choi isomorphism \cite{harrow2013testing}.
The tasks of testing Pauli matrices and Clifford gates are also studied in \cite{montanaro2010quantum} and \cite{low2009learning,wang2011property} respectively.

The structure of the present paper is as follows.
Section \ref{sec2} gives a brief introduction to group representation theory which is a basic method for our analysis.
Section \ref{sec3} studies the certification of the membership of an unknown state in a known set.
Section \ref{sec5} studies the certification of the equality of two unitary quantum channels.
Finally Section \ref{sec6} gives a discussion.




 




\section{Review on representation theory}\label{sec2}
In this section we give a brief introduction to important results in group representation theory and fix some notation, which will be used later.
For more details on representation theory, we refer to~\cite{fulton2013representation,goodman2009symmetry,hall2015lie}.
Consider the following Schur-Weyl duality
\Eqn{
	(\Cb^d)^{\ot n} = \Oplus_{\ld\vdash(n,d)} \Hc_\ld\ot\Kc_\ld \,,
}
where $\Hc_\ld$ is the irreducible representation (irrep) of unitary group $\x{U}(d)$ corresponding to partition $\ld$ and $\Kc$ is the corresponding irrep of the symmetric group $S_n$.
Here $\ld\vdash(n,d)$, also written $\ld\in\x{Par}(n,d)$, means that $\ld$ is a partition of $n$ with length at most $d$.

The dimension of $\Kc_\ld$, for partition $\ld$ of $n$, is given by
\Eq{\label{fjoiwhoggefho}
\dim \Kc_\ld = \f{n!}{\widetilde{\ld}_1!\cdots\widetilde{\ld}_d!}\prod_{1\le i<j\le d}\big(\widetilde{\ld}_i-\widetilde{\ld}_j\big) \,,
}
where $\widetilde\ld_i:=\ld_i+d-i$.
It can be bounded by \cite{hayashi2002exponents}
\Eq{
\binom{n}{\ld}(n+d)^{-d(d-1)/2} \le \dim V_\ld^S \le \binom{n}{\ld}
}
and furthermore
\Eq{\label{vmognreogbjgt2}
	\exp(nH(\bar \ld)) (n+d)^{-d(d+1)/2} \le\dim V_\ld^S \le \exp(nH(\bar \ld)) \,,
}
where $\bar\ld:=\ld/n$.

The character $\chi_\ld^L$ for the irrep $\Hc_\ld$ is
\Eq{\label{owdhbeobhro}
\chi_\ld^L(\x{diag}(x_1,\dots,x_d))=\f{\det(x_i^{\ld_j+d-j})_{i,j=1}^d}{\det (x_i^{d-j})_{i,j=1}^d} \,.
}
The dimension of $\Hc_\ld$ is given by
\Eq{\label{vmognreogbjgt}
\dim \Hc_\ld = \prod_{1\le i<j \le d}\f{\ld_i-\ld_j+j-i}{j-i} \,,
}
which is bounded above by $(n+1)^{d(d-1)/2}$ \cite{christandl2006spectra}.

\section{Testing membership of a finite set of states}\label{sec3}



In the study of testing whether an unknown state is close to or far from a given state and whether two unknown states are close to each other, B\u{a}descu {\it et al.}\ \cite{buadescu2019quantum} used the Chebyshev inequality to bound the deviation of a random variable from its expectation so that the completeness and soundness conditions can be satisfied.
To be specific, for a real-valued random variable $X$ and a scalar $0<\gm<\f{1}{2}$, consider a tester which reports $\E X\le (1-2\gm)\tht$ when $X\le(1-\gm)\tht$ is observed and reports $\E X\ge \tht$ when $X>(1-\gm)\tht$ is observed.
When $\E X\le (1-2\gm)\tht$, the probability that the tester reports correctly is
\Eq{\label{vnjgrwhteo}
\Pr(X\le(1-\gm)\tht)\ge 1-\f{\x{Var} X}{((1-\gm)\tht-\E X)^2}\ge 1-\f{\x{Var} X}{(\gm\tht)^2} \,.
}
When $\E X\ge\tht$, the probability that the tester reports correctly is
\Eq{\label{vnjgrwhteo2}
\Pr(X>(1-\gm)\tht)\ge1-\f{\x{Var} X}{(\E X-(1-\gm)\tht)^2}\ge 1-\f{\x{Var} X}{(\gm\tht)^2} \,.
}
If $\x{Var} X$ is small enough as compared with $\gm\tht$, the tester is eligible.
In our settings, let $X$ denote the random variable of measurement outcome of $M$ on a state $\varrho$.
Obviously the expectation of $X$ is $\E X=\tr(\varrho M)$ and the variance of $X$ is $\x{Var} X=\tr(\varrho M^2)-(\tr \varrho M)^2$.
A quantum algorithm was proposed in \cite{buadescu2019quantum} to test whether an unknown state is close to or far from a fixed state using $O(\epsl^{-2})$ copies of the state.


In order to test whether an unknown state is contained in or far from a given finite set of states, we need a revised tester.
For the random variable $Y=\min\{X_1,\dots,X_m\}$ where $X_i$'s are independent random variables, consider the tester which accepts (i.e., reports $\min_i\E X_i\le(1-2\gm)\tht$) when $\min_i X_i\le(1-\gm)\tht$ and rejects (i.e., reports $\min_i\E X_i\ge\tht$) otherwise.

When $\min \E X_i\le(1-2\gm)\tht$, we have
\Als{\label{ohguoreds}
\Pr(\min X_i\le(1-\gm)\tht) &=1-\prod\nolimits_i \Pr(X_i>(1-\gm)\tht) \\
&\ge 1-\Pr(X_k>(1-\gm)\tht) \\
&\ge 1-\f{\x{Var} X_k}{(\gm\tht)^2} \,,
}
where $\E X_k=\min_i\E X_i$.

When $\min \E X_i\ge\tht$, we have
\Als{\label{ohguoreds2}
\Pr(\min X_i>(1-\gm)\tht) &= \prod\nolimits_i \Pr(X_i>(1-\gm)\tht) \\
& \ge \prod\nolimits_i \Big( 1-\f{\x{Var} X_i}{(\E X_i-(1-\gm)\tht)^2}\Big) \,.
}

Using the above bounds, following the idea in \cite{buadescu2019quantum}, we have

\Lem{
	Given a finite set $\Pc$ of states and an unknown state $\rho$, there exists a tester that distinguishes whether $\rho\in\Pc$ or $\min_{\sg\in\Pc} \|\rho-\sg\|_2\ge \epsl$ using $O(\epsl^{-2} |\Pc|)$ copies of the unknown state.
}
\begin{proof}
Take $\gm=\f{1}{2}$, $\tht=\epsl^2$ and denote $m=|\Pc|$.
Using Eq.~\eqref{ohguoreds} and the estimate of variance of $X_k$, $O(\epsl^{-2})$ copies of $\rho$ suffice to ensure that $\Pr(\min X_i\le(1-\gm)\tht)\ge 2/3$.
For Eq.~\eqref{ohguoreds2}, $\Pr(\min X_i>(1-\gm)\tht)\ge \big(1-c\epsl^{-4}(\f{1}{n^2}+\f{\epsl^2}{n})\big)^m$ for some constant $c$ and large $n$.
So $n$ should satisfy that $1-c\epsl^{-4}(\f{1}{n^2}+\f{\epsl^2}{n})\ge (\f{2}{3})^{1/m}$.
In order for this inequality to hold, take $n=c\epsl^{-2} \big(1-(\f{2}{3})^{1/m}\big)^{-1}=O(\epsl^{-2}m)$.
\end{proof}

We now apply the exponential Markov inequality to bound the deviation of $X$ from its expectation, yielding an alternative sample complexity for testing the property of a finite set of pure states.
When $\E X\le (1-2\gm)\tht$, the tester accepts with probability
\Eq{\label{cnowehbhdaoew}
\Pr(X\le (1-\gm)\tht)=1-\Pr(X> (1-\gm)\tht)\ge 1-\f{\E\ee^{sX}}{\ee^{s(1-\gm)\tht}} \,,
}
for any $s>0$.
When $\E X\ge\tht$, then the tester rejects with probability
\Eq{\label{cnowehbhdaoew2}
\Pr(X>(1-\gm)\tht)=1-\Pr(X\le(1-\gm)\tht)\ge1-\f{\E\ee^{sX}}{\ee^{s(1-\gm)\tht}} \,,
}
for any $s<0$.

\Thm{
	Given a finite set $\Pc$ of pure states and an unknown pure state $\psi$, there exists a tester with perfect completeness that distinguishes whether $\psi\in\Pc$ or $\x{D}(\psi,\Pc)\ge \epsl$ using $O(\epsl^{-4} \ln |\Pc|)$ copies of the unknown state.
}
\begin{proof}
For any $\phv\in\Pc$, choose an orthonormal basis such that $\phv=\x{diag}(1,0,\dots,0)$ in this basis, and denote the diagonal elements of $\psi$ by $x_1,x_2,\dots,x_d$ in this basis.
Let $\Pi_t$ denote the projector onto the subspace of $(\Cb^d)^{\ot n}$ spanned by states of type $t$.
Consider the observable $M=\sum_{t\in\x{Type}(n,d)}\f{t_1}{n}\Pi_t$ which was first used in \cite{buadescu2019quantum}.
Let $X$ denote the measurement outcome of $M$ on $\psi^{\ot n}$, then $\E X=\tr(\psi^{\ot n}M)=\brak{\psi,\phv}=x_1$.
The moment-generating function of $X$ is
\Aln{
\E\ee^{sX}=\tr(\psi^{\ot n}\ee^{sM}) &=\sum_{t_1+\cdots+t_d=n}\ee^{st_1/n}\tr(\psi^{\ot n}\Pi_t) \\
&=\sum_{t_1=0}^n\ee^{st_1/n}\sum_{t_2+\cdots+t_d=n-t_1}\tr(\psi^{\ot n}\Pi_t) \\
&=\sum_{t_1=0}^n\ee^{st_1/n}\sum_{t_2+\cdots+t_d=n-t_1} \binom{n}{t} x_1^{t_1}\cdots x_d^{t_d} \\
&=\sum_{t_1=0}^n\ee^{st_1/n} \binom{n}{t_1} x_1^{t_1} (1-x_1)^{n-t_1} \\
&=\big(1+(\ee^{s/n}-1)x_1\big)^n \,,
}
where $\binom{n}{t}:=\f{n!}{t_1!\cdots t_d!}$ and $\binom{n}{t_1}:=\f{n!}{t_1!(n-t_1)!}$.


When $\E X=x_1\le 1-\epsl^2$, take $s=2n\epsl^2$.
Since $x_1=\brak{\psi,\phv}\le 1-\epsl^2$, we have
\Aln{
	\E\ee^{sX} &\le \big( 1+(\ee^{2\epsl^2}-1)(1-\epsl^2) \big)^n \\
	&= ( 1+2\epsl^2+O(\epsl^6) )^n \,.
}

Using Eqs.~\eqref{cnowehbhdaoew} and~\eqref{cnowehbhdaoew2} with $\gm=\epsl^2/2$ and $\tht=1$, it follows that
\Aln{
\Pr(X\le 1-\epsl^2/2 ) &\ge 1-\f{ \E\ee^{sX} }{\ee^{s(1-\epsl^2/2)}} \\
& \ge 1-\f{ ( 1+2\epsl^2+O(\epsl^6) )^n }{ \ee^{n(2\epsl^2-\epsl^4)} } \\
&=1-(1-2\epsl^4+O(\epsl^6))^n \,.
}

When $\E X=x_1=1$, the measurement outcome $X$ of $M$ is $1$ with certainty.

Now we consider the task of testing whether an unknown state $\psi$ is contained in or far from the set $\Pc$.
For each $\phv\in\Pc$, one can define an corresponding observable $M_\phv$.
Measure the observable $M_\phv$ on the state $\psi^{\ot n}$, and denote the outcome by $X_\phv$.
Consider the tester that reports $\x{D}(\psi,\Pc)\ge\epsl$ i.e., $\max_\phv \E X_\phv\le 1-\epsl^2$ when $\max_\phv X_\phv\le 1-\f{\epsl^2}{2}$, and reports $\psi\in\Pc$ otherwise.
For the soundness condition, in order for
\Eqn{
\Pr\big(\max\nolimits_\phv X_\phv\le 1-\epsl^2/2\big)=\Pi_\phv \Pr(X_\phv\le1-\epsl^2/2)\ge \big(1-(1-2\epsl^4+O(\epsl^6))^n\big)^{|\Pc|}\ge \f{2}{3}
}
to hold, it suffices to take
\Eqn{
n=\f{ \ln\big(1-(\f{2}{3})^{1/|\Pc|}\big) }{ \ln( 1-2\epsl^4+O(\epsl^6) ) } =O(\epsl^{-4}\ln|\Pc|) \,.
}

As for the completeness condition, when $\psi\in\Pc$, $\max\nolimits_\phv X_\phv=1$.
Thus this tester has perfect completeness.
Therefore using $O(\epsl^{-4}\ln|\Pc|)$ copies of $\psi$ the tester fulfills both completeness and soundness conditions.
\end{proof}

Denote $\dt:=\{\x{D}(\phv_1,\phv_2):\phv_1\ne\phv_2\in\Pc\}$.
It was shown in \cite{wang2011property} that $O(\max\{\epsl^{-2},\dt^{-2}\}\ln|\Pc|)$ copies of states suffice to distinguish whether $\psi\in\Pc$ or $\x{D}(\psi,\Pc)\ge\epsl$.
When $\dt=o(\epsl^2)$, our method is more sample-efficient than that in \cite{wang2011property}.
The method based on Schur-Weyl decomposition used in this chapter may be useful in the certification in distributed scenarios.



\section{Testing equality of unitary quantum channels}\label{sec5}


Following \cite{montanaro2016survey}, define the distance between two unitary matrices $U,V$ of order $d$ as
\Eq{\label{defjfogbben}
\x{dist}(U,V)=\sqrt{1-\Big|\f{1}{d}\brak{U,V}\Big|^2} \,,
}
where $\brak{U,V}:=\tr(U\dg V)$.
It can be seen that the distance of two unitaries is at most 1.

Notice that the normalized inner product of unitaries is equal to the inner product of their Choi states, i.e.,
\Eq{
\f{1}{d}\brak{ U,V }=\bra{\phi^\x{m}} (U\dg\ot\1)(V\ot\1) \ket{\phi^\x{m}} \,.
}
Here and in the following we use $\ket{\phi^\x{m}}$ to denote the (normalized) maximally entangled state.
Consider the following Schur-Weyl decomposition
\Eqn{
	(\Cb^d)^{\ot n} = \Oplus_{\ld\vdash(n,d)} \Hc_\ld\ot\Kc_\ld \,,
}
where $\Hc_\ld$ is the irrep of $\Ux(d)$ and $\Kc_\ld$ is its corresponding multiplicity space.
The entanglement between the representation space and multiplicity space was used to estimate the group transformation \cite{chiribella2005optimal}.
It turns out that it can be also used in the certification of unitaries as shown in the following.

\Thm{\label{thm:vnewrhod}
	Given access to an unknown single-qubit unitary $U$, there exists a tester with perfect completeness that distinguishes whether $U$ is equal to a fixed and known $V$ up to a phase or $\x{dist}(U,V)\ge\epsl$ with $O(\epsl^{-1})$ uses of $U$, without using any ancilla system.
}
\begin{proof}
By Schur-Weyl duality, $(\Cb^2)^{\ot n}=\Oplus_j \Hc_j\ot\Kc_j$ where $\Hc_j$ is the irrep of $\Ux(2)$ of dimension $2j+1$ and $\Kc_j$ is the corresponding irrep of $S_n$.
In this decomposition we write $U^{\ot n}=\Oplus_j U_j\ot\1_j$ for any $U\in\Ux(2)$ since $U^{\ot n}$ acts as identity operator on each $\Kc_j$.

Notice that $\Hc_{\f{n}{2}-1}$ and $\Kc_{\f{n}{2}-1}$ have the same dimension $n-1$.
Let $\ket{\phi^\x{m}}$ be the maximally entangled state in $\Hc_{\f{n}{2}-1}\ot\Kc_{\f{n}{2}-1}$, i.e., $\ket{\phi^\x{m}}=\f{1}{\sqrt{n-1}}\sum_{i=1}^{n-1}\ket{\alp_i}\ket{\bt_i}$ where $\{\ket{\alp_i}\}$ and $\{\ket{\bt_i}\}$ are orthonormal bases of $\Hc_{\f{n}{2}-1}$ and $\Kc_{\f{n}{2}-1}$ respectively.

Apply the unitary $U^{\ot n}$ to $\ket{\phi^\x{m}}$, and then perform the POVM $\{ \phi^\x{m}_V,\1-\phi^\x{m}_V \}$ where $\ket{\phi^\x{m}_V}:=(V_{\f{n}{2}-1}\ot\1_{\f{n}{2}-1})\ket{\phi^\x{m}}$.
The tester reports that $U$ equals $V$ up to a phase if the first outcome occurs, and reports $\x{dist}(U,V)\ge\epsl$ otherwise.
Obviously the tester has perfect completeness irrespective of the choice of $n$.
Next step is to derive the requirement for $n$ to ensure the soundness condition $\brak{ U^{\ot n}\phi^\x{m}U^{\dagger,\ot n},\phi^\x{m}_V }\le1/3$ when $\x{dist}(U,V)\ge\epsl$.

Denote the eigenvalues of $U\dg V$ by $\ee^{\ii\alp}$ and $\ee^{\ii\bt}$ for $\alp,\bt\in[0,2\pi)$.
When $\x{dist}(U,V)\ge\epsl$, then by the definition~\eqref{defjfogbben}, $|\brak{U,V}|^2\le4(1-\epsl^2)$, i.e., $|\ee^{\ii\alp}+\ee^{\ii\bt}|^2\le4(1-\epsl^2)$.
It follows that $|\sin\f{1}{2}(\alp-\bt)|\ge\epsl$.
We thus have
\Al{\label{ceguorbgwogb}
\sqrt{ \brak{ U^{\ot n}\phi^\x{m}U^{\dagger,\ot n},\phi^\x{m}_V } } &= \Big|\f{1}{n-1}\big\langle U_{\f{n}{2}-1},V_{\f{n}{2}-1} \big\rangle\Big| \notag \\
&=\Big|\f{1}{n-1} \f{\ee^{\ii(n\alp+\bt)}-\ee^{\ii(\alp+n\bt)}}{\ee^{\ii\alp}-\ee^{\ii\bt}} \Big|\notag \\
&=\Big|\f{1}{n-1} \f{\sin\f{n-1}{2}(\alp-\bt)}{\sin\f{1}{2}(\alp-\bt)} \Big| \notag \\
&\le \Big|\f{1}{n-1} \f{1}{\sin\f{1}{2}(\alp-\bt)} \Big| \notag\\
&\le \f{1}{(n-1)\epsl} \,,
}
where the second equality used the Weyl character formula~\eqref{owdhbeobhro} for irrep $\Hc_{\f{n}{2}-1}$.
We now take $n=\f{\sqrt3}{\epsl}+1=O(\epsl^{-1})$ so that the soundness condition $\brak{ U^{\ot n}\phi^\x{m}U^{\dagger,\ot n},\phi^\x{m}_V }\le1/3$ for $\x{dist}(U,V)\ge\epsl$ is satisfied.
\end{proof}


When both single-qubit unitaries are unknown, we have the following.
\Thm{\label{thm:vnewrhod2}
	Given access to two unknown single-qubit unitaries $U$ and $V$, there exists a tester with perfect completeness that distinguishes whether $U$ equals $V$ up to a phase or $\x{dist}(U,V)\ge\epsl$ with $O(\epsl^{-1})$ uses of $U$ and $V$, without using any ancilla system.
}
\begin{proof}
Similar to the proof of Theorem~\ref{thm:vnewrhod}, we first decompose the space as $(\Cb^2)^{\ot n}=\Oplus_j \Hc_j\ot\Kc_j$ where $\Hc_j$ and $\Kc_j$ are irreps of $\Ux(2)$ and $S_n$ respectively.
Denote by $\ket{\phi^\x{m}}$ the maximally entangled state in $\Hc_{\f{n}{2}-1}\ot\Kc_{\f{n}{2}-1}$.
Then we apply the swap test to $U^{\ot n}\ket{\phi^\x{m}}$ and $V^{\ot n}\ket{\phi^\x{m}}$.
Repeat the above procedure, and report that the two unitaries are equal up to a phase if and only if the swap test accepts twice.
When $U$ and $V$ are equal, the tester reports correctly with certain.
When $\x{dist}(U,V)\ge\epsl$, since $\big|\brak{ \phi^\x{m}|U^{\dagger,\ot n}V^{\ot n}|\phi^\x{m} }\big|\le \f{1}{(n-1)\epsl}$ by~\eqref{ceguorbgwogb}, the tester reports incorrectly with probability less than $\big(\f{1}{2}(1+\f{1}{(n-1)^2\epsl^2})\big)^2$, which is in turn less than $1/3$ if we take $n=\f{3}{\epsl}+1$.
\end{proof}

In order to deal with the higher dimensional case, an upper bound on $|\tr U_\ld|/d_\ld$ is needed given $|\tr U|/d\le 1-\epsl^2$, where $U_\ld$ is representation matrix of $U$ on irrep $\Hc_\ld$ for appropriate $\ld\in\x{Par}(n)$ and $d_\ld$ is the dimension of $\Hc_\ld$.

\Lem{\label{lemma-blowup}
	Let $\ld=(d-1,d-2,\dots,0) n/\binom{d}{2}\in\x{Par}(n,d)$ and $s=n/\binom{d}{2}+1$.
	If $\f{1}{d}|\brak{U,V}|\le1-\epsl^2$ for $U,V\in\Ux(d)$, then
	\Eqn{
	\f{1}{\dim\Hc_\ld}|\brak{U_\ld,V_\ld}| \le \Big(\f{2}{s\epsl}\Big)^m
	}
	for some positive $m$.
}
\begin{proof}
The dimension of the irrep $\Hc_\ld$ is
\Eq{\label{dimvowehgwbf}
	\dim\Hc_\ld = s^{d(d-1)/2} \,.
}
The character for $\Hc_\ld$ is
\Eqn{
	\chi_\ld^L(\x{diag}(x_1,\dots,x_d)) = \f{ \prod_{1\le j<k \le d}(x_k^s-x_j^s) }{ \prod_{1\le j<k \le d}(x_k-x_j) } \,.
}

Since $\s{R}_\ld:U\mapsto U_\ld$ is a unitary representation of $\Ux(d)$, we have $U_\ld V_\ld=(UV)_\ld$ and $(U_\ld)\dg=(U\dg)_\ld$.
Thus $\brak{U_\ld,V_\ld}=\brak{(U\dg V)_\ld,\1}$.
It suffices to consider the case $V=\1$.

Consider a unitary $U$ having eigenvalues $\ee^{\ii\tht_k}$ with $0\le \tht_k<2\pi$ for each $k\in[d]$.
We have
\Al{\label{vjengornfkdne}
	|\tr U|^2 = \bigg|\sum_{k=1}^d \ee^{\ii\tht_k}\bigg|^2 &=  d+2\sum\nolimits_{1\le j<k\le d}\cos(\tht_k-\tht_j) \nonumber\\
	&= d^2-4\sum\nolimits_{1\le j<k\le d}\sin^2\f{1}{2}(\tht_k-\tht_j)  \,,
}
while
\Al{\label{vjenfkdnewiojrnwrg}
| \tr U_\ld | &= \prod_{1\le j<k \le d} \bigg| \f{ \ee^{\ii s\tht_k}-\ee^{\ii s\tht_j} }{ \ee^{\ii \tht_k}-\ee^{\ii \tht_j} } \bigg| \nonumber \\ 
&= \prod_{1\le j<k \le d}\bigg| \f{ \sin\f{s}{2}(\tht_k-\tht_j) }{ \sin\f{1}{2}(\tht_k-\tht_j) } \bigg| \,.
}

Before proceeding with the proof we now show that
\Eq{\label{wohvoeaufcvd}
\Big|\f{\sin sx}{\sin x}\Big|\le s
}
holds for any odd positive integer $s$ and any $|x|\le \pi $.
Indeed, it suffices to consider the case $0\le \sin x\le\f{1}{s}$, and~\eqref{wohvoeaufcvd} follows by noticing that the function $\f{\sin sx}{s\sin x}$ is even and is symmetric about the line $x=\f{\pi}{2}$.
When $0\le \sin x\le\f{1}{s}$, since $\sin x\ge \f{2}{\pi}x$, we have $sx\le\f{\pi}{2}$.
It follows that for any $s\ge1$, we have $\cos x\ge\cos sx$, and thus $s\sin x\ge\sin sx$, completing the proof of~\eqref{wohvoeaufcvd}.

As for the soundness condition, when $\f{1}{d}|\tr U|\le 1-\epsl^2$, by~\eqref{vjengornfkdne} we have
\Eqn{
	\sum_{1\le j<k\le d}\sin^2\f{1}{2}(\tht_k-\tht_j)\ge \f{1}{4}d^2(2\epsl^2-\epsl^4) \,.
}

It follows that there exists $(j,k)$ satisfying
\Eqn{
\sin^2\f{1}{2}(\tht_k-\tht_j)\ge \f{d(2\epsl^2-\epsl^4)}{2(d-1)} \,,
}
and let $m$ be the number of such pairs in totally $\binom{d}{2}$ pairs.
For any such pair,
\Eq{\label{vekrghoer}
\f{ |\sin\f{s}{2}(\tht_k-\tht_j)| }{ |\sin\f{1}{2}(\tht_k-\tht_j)| } \le \f{ 1 }{ |\sin\f{1}{2}(\tht_k-\tht_j)| } \le \sqrt{ \f{2(d-1)}{d(2\epsl^2-\epsl^4)} }\le \f{2}{\epsl}
}
for $\epsl\le1$.
On the other hand, there are $\binom{d}{2}-m$ pairs of $(j,k)$ satisfying $\sin^2\f{1}{2}(\tht_k-\tht_j) < \f{d(2\epsl^2-\epsl^4)}{2(d-1)}$.
For any such pair, since $|\f{1}{2}(\tht_k-\tht_j)| \le\pi$, by~\eqref{wohvoeaufcvd} we have
\Eq{\label{vekrghoer222}
\f{ |\sin\f{s}{2}(\tht_k-\tht_j)| }{ |\sin\f{1}{2}(\tht_k-\tht_j)| }\le s \,.
}

Therefore, combining~\eqref{dimvowehgwbf},~\eqref{vjenfkdnewiojrnwrg},~\eqref{vekrghoer} and~\eqref{vekrghoer222},
\[\f{1}{\dim\Hc_\ld}|\tr U_\ld|\le s^{-\binom{d}{2}} (2/\epsl)^m s^{-m+\binom{d}{2}} =\Big(\f{2}{s\epsl}\Big)^m \,.
\]
\end{proof}


\Thm{\label{dgwegnkwebngjbw}
	Given access to an unknown unitary $U\in\Ux(d)$ and a known or unknown unitary $V\in\Ux(d)$, there exists a tester with perfect completeness that distinguishes whether $\x{dist}(U,V)=0$ or $\x{dist}(U,V)\ge\epsl$ with $O(d^2/\epsl)$ uses of unitaries.
}


\begin{proof}
Consider the partition $\ld=(d-1,d-2,\dots,0) n/\binom{d}{2}$, and denote $s:=n/\binom{d}{2}+1$.
The proof follows similar approach used in Theorems~\ref{thm:vnewrhod} and~\ref{thm:vnewrhod2}.

By noticing that $\ld/n$ is independent of $n$, it follows from~\eqref{vmognreogbjgt2} and~\eqref{vmognreogbjgt} that when $\epsl$ is so small that $n$ is much larger than $d$, the dimension of $\Kc_\ld$ is exponential in $n$ while the dimension of $\Hc_\ld$ is polynomial in $n$, thus $\dim\Kc_\ld$ is larger than $\dim\Hc_\ld$.

Denote by $\ket{\phi^\x{m}}$ the maximally entangled state in $\Hc_\ld\ot\Kc_\ld$, and denote $\ket{\phi_U^\x{m}}:=U^{\ot n}\ket{\phi^\x{m}}$ and $\ket{\phi_V^\x{m}}=V^{\ot n}\ket{\phi^\x{m}}$.
When $U$ is unknown and $V$ is given, perform a measurement $\{\phi_V^\x{m},\1-\phi_V^\x{m}\}$ on $\phi_U^\x{m}$, and repeat, and accept iff the first outcome occurs.
When $U$ and $V$ are both unknown, perform a swap test for $\phi_U^\x{m}$ and $\phi_V^\x{m}$, and repeat.

The completeness condition is obvious.
When $\x{dist}(U,V)\ge\epsl$, using Lemma~\ref{lemma-blowup}, the soundness condition is satisfied by taking $s$ to be the smallest odd integer larger than $6/\epsl$, for which $n\le \binom{d}{2}(\f{6}{\epsl}+1)=O(d^2/\epsl)$.
\end{proof}


For the asymptotic case where the value of $\epsl$ is small, $n$ will be so large that the irrep $\Hc_\ld$ has smaller dimension than its multiplicity space $\Kc_\ld$ does.
Thus there exists a maximally entangled state $\phi^\x{m}$ on $\Hc_\ld\ot\Hc_\ld$ which is a subspace of $\Hc_\ld\ot\Kc_\ld$.
If $\epsl$ is not small enough and thus $n$ is not large enough, one can introduce a reference system of dimension $\f{\dim\Hc_\ld}{\dim\Kc_\ld}$ to make them have equal dimension \cite{chiribella2005optimal}.
The dimension of the ancilla system, however, has been exponentially decreased compared with the usual method using Choi states directly.
Since it is common to deal with low-dimensional quantum systems in quantum computing under current technology, our method exhibits advantage in practice.

\section{Discussion}\label{sec6}

We have proposed and analyzed in this work efficient algorithms for certification of quantum state set and quantum unitary.

For the certification of identity of qudit unitaries, we have considered the irrep corresponding partition $\ld=(d-1,d-2,\dots,0)n/\binom{d}{2}$.
One issue that would be worth investigating is whether this irrep is optimal compared with other irreps.

Using Theorem~\ref{dgwegnkwebngjbw}, one can also efficiently test whether one unitary is identity, and whether two unitaries are inverse to each other.
It is known that many properties of quantum channel can be reduced to testing properties of quantum state via the Choi-Jamio{\l}kowski isomorphism, but by employing this approach one usually needs to use quantum channels too many times and also to introduce extra ancilla system.
We used group representation to test properties of unitaries more sample-efficiently without using ancilla system.
The method based on group representation theory may be useful in testing other properties of quantum states and unitaries, which deserves further study.
Our approach may be promising in the property testing of noisy quantum operation and measurement to achieve better performance, which remains further study.



\section*{Acknowledgements}
This work was supported by the School of Computer Science and Technology, University of Science and Technology of China (Grant No.\ KY0110009999).

\bibliographystyle{plain} \bibliography{bib}


\end{document}